\documentclass[prl,aps,showpacs,twocolumn,floatfix]{revtex4}
\usepackage{epsfig} \usepackage{graphics} \usepackage{bm}
\usepackage{amssymb}
\usepackage{graphicx}
\begin{document}

\preprint{Lebed-Sepper-PRL}

\title{Quantum Limit in a Magnetic Field for Triplet Superconductivity
in a Quasi-One-Dimensional Conductor}

\author{A.G. Lebed$^*$}
\author{O. Sepper}

\affiliation{Department of Physics, University of Arizona, 1118 E.
4-th Street, Tucson, AZ 85721, USA}

\begin{abstract}
We theoretically consider the upper critical magnetic field,
perpendicular to a conducting axis in a triplet
quasi-one-dimensional superconductor. In particular, we
demonstrate that, at high magnetic fields, the orbital effects
against superconductivity in a magnetic field are reversible and,
therefore, superconductivity can restore. It is important that the
above mentioned quantum limit can be achieved in presumably
triplet quasi-one-dimensional superconductor
Li$_{0.9}$Mo$_6$O$_{17}$ [J.-F. Mercure et al., Phys. Rev. Lett.
{\bf 108}, 187003 (2012)] at laboratory available pulsed magnetic
fields of the order of $H = 500-700 \ T$.
\end{abstract}


\maketitle

High magnetic field properties of quasi-one-dimensional (Q1D)
conductors and superconductors have been intensively studied since
the discovery of the Field-Induced Spin-Density-Wave (FISDW)
cascade of phase transitions [1-4]. Note that the FISDW
phenomenon, experimentally discovered in the (TMTSF)$_2$X
compounds [1,2], where X=ClO$_4$ and X=PF$_6$, was historically
the first one which was successfully explained in terms of
quasi-classical $3D \rightarrow 2D$ dimensional crossover in high
magnetic fields [3,5-7]. At present, it has been established that
different kinds of quasi-classical dimensional crossovers in a
magnetic field are responsible for such unusual phenomena in
layered Q1D conductors as the Field-Induced Charge-Density-Wave
(FICDW) phase transitions [3,5,8,9], Danner-Kang-Chaikin oscillations
[3,10], Lebed Magic Angles [3,11,12], and Lee-Naughton-Lebed
oscillations [3,13-16]. Note that a characteristic property of the
quasi-classical dimensional crossovers is that the typical "sizes"
of electron trajectories in a magnetic field are much lager than
the inter-plane or inter-chain distances in layered Q1D
conductors.

On the other hand, a different type of dimensional crossovers in a
magnetic field - the so-called quantum dimensional crossover [3] -
was suggested in Ref.[17] to describe magnetic properties of a
superconducting phase. More specifically, it was shown [17-22]
that, at high enough magnetic fields, the typical "sizes" of
electron trajectories become of the order of inter-plane distances
and superconductivity can restore as a pure 2D phase. Note that
the above mentioned conclusion is valid only for some triplet
superconducting phases which are not sensitive to the Pauli
paramagnetic effects in a magnetic field. Due to this reason, Q1D
superconductors (TMTSF)$_2$X were considered for many years to be
the best candidates for this Reentrant Superconductivity (RS)
phenomenon, since triplet superconducting pairing was suggested
[23,24] to exist in these materials. Recently, it has been shown
[25,22,26] that d-wave singlet superconducting phase is more
likely to exist in the (TMTSF)$_2$ClO$_4$, therefore, the RS
phenomenon experimentally reveals itself in this compound only as
the hidden RS phase [22]. As to the superconductor
(TMTSF)$_2$PF$_6$, the existing experimental data about nature of
superconducting pairing in this compound are still controversial
[3]. In this difficult situation, it is important that a new
strong candidate for triplet superconducting pairing - the Q1D
superconductor Li$_{0.9}$Mo$_6$O$_{17}$ - has been recently
proposed [27-29]. In particular, it has been shown in Refs.[28,29]
that a quantitative description by triplet scenario of
superconductivity, with a magnetic field applied along the
conducting axis, can account for the experimental field that
exceeds the so-called Clogston-Chandrasekhar paramagnetic limit
[30] by five times [27].

The goal of our Letter is to show theoretically that triplet
superconductivity scenario can be tested in the
Li$_{0.9}$Mo$_6$O$_{17}$ superconductor in ultra-high magnetic
fields of the order of $H \simeq 500-700 \ T$, where triplet
superconductivity is shown to restore with transition temperature
$T^*_c \simeq 0.75 \ T_c \simeq 1.4-1.7 \ K$. Note that the
suggested effect is different from the RS phenomenon [17-22] since
we consider a magnetic field, which has non-zero out-off
conducting plane component. Therefore, at high magnetic fields,
$3D \rightarrow 2D$ crossover [17,3] does not happen. Instead,
quantum $3D \rightarrow 1D$ dimensional crossover happens which,
to the best of our knowledge, has not been considered and studied
so far. We call such crossover $Q1D \rightarrow 1D$ quantum limit
(QL) [31]. [Note that below we apply the Fermi liquid approach to
the Q1D transition-metal oxide Li$_{0.9}$Mo$_6$O$_{17}$. Validity
of the Fermi liquid picture as well as Q1D nature of electron
spectrum in this conductor have been firmly established in
Refs.[27-29].]

First, let us demonstrate the suggested in the Letter QL
superconductivity phenomenon using qualitative arguments. Below,
we consider electron spectrum of a Q1D conductor in a
tight-binding approximation,
\begin{equation}
\epsilon({\bf p})= - 2t_x \cos(p_x a_x) - 2 t_y \cos(p_y a_y) - 2t_z
\cos (p_z a_z),
\end{equation}
where $t_x \gg t_y > t_z$ are overlap integrals of the electron
wave functions along ${\bf x}$, ${\bf y}$, and ${\bf z}$
crystallographic axes, respectively. Since $t_x \gg t_y , t_z$,
the electron spectrum (1) corresponds to two slightly deformed
pieces of the Fermi surface (FS) and can be linearized near $p_x
\simeq \pm p_F$,
\begin{equation}
\epsilon({\bf p})= \pm v_F(p_x \mp p_F) - 2 t_y \cos(p_y a_y) -
2t_z \cos (p_z a_z),
\end{equation}
(see Fig.1) where $p_F$ and $v_F$ are the Fermi momentum and Fermi
velocity, respectively. In a magnetic field, perpendicular to the conducting
axis,
\begin{equation}
{\bf H} = (0,\cos \alpha,\sin \alpha)H, \ {\bf A} = (0,-\sin
\alpha ,\cos \alpha ) Hx,
\end{equation}
(see Fig.2) it is possible to write quasi-classical equations of electron
motion,
\begin{equation}
\frac{d{\bf p}}{dt}= \bigl(\frac{e}{c}\bigl) [{\bf v({\bf p})} \times {\bf
H}],
\end{equation}
 in the following way:
\begin{equation} \frac{d(p_z a_z)}{dt}=\omega_z(\alpha)t, \ \
\frac{d(p_y a_y)}{dt}= - \omega_y(\alpha)t,
\end{equation}
where
\begin{equation}
\omega_y(\alpha)=e v_F a_y H \sin \alpha /c, \ \
\omega_z(\alpha)=e v_F a_z H \cos \alpha /c.
\end{equation}
Since electron velocity components can be expressed as functions
of time,
\begin{eqnarray}
v_y(p_y)=\partial \epsilon({\bf p})/\partial p_y = -2t_ya_y
\sin[\omega_y(\alpha)t] \ , \nonumber\\
v_z(p_z)=\partial \epsilon({\bf p})/\partial p_z = 2t_za_z
\sin[\omega_z(\alpha)t] \ ,
\end{eqnarray}
their trajectories in a real space are described by the following
equations:
\begin{eqnarray}
y = l_y(\alpha)a_y \cos [\omega_y(\alpha)t], \ l_y(\alpha)=
2t_y/\omega_y(\alpha),
\nonumber\\
z = - l_z(\alpha)a_z \cos [\omega_z(\alpha)t], \ l_z(\alpha)=
2t_z/\omega_z(\alpha) \ .
\end{eqnarray}

As directly follows from Eq.(8), electron motion in a real space
in the magnetic field (3) is free along the conducting axes and
periodic and restricted perpendicular to the axes. If
the magnetic field is strong enough,
\begin{equation}
H \gg H^* = \max \biggl\{ \frac{2t_y c}{ev_Fa_y \sin \alpha} \ ,
\frac{2t_z c}{ev_Fa_z \cos \alpha} \biggl\} \ ,
\end{equation}
then electron motion in perpendicular directions becomes localized
on the conducting axes. This fact is directly seen from
Eqs.(8),(9) since the characteristic "sizes" of electron
trajectories, $l_y(\alpha) \ a_y$ and $l_z(\alpha) \ a_z$, become
less than the corresponding inter-chain distances, $a_y$ and
$a_z$. In this case, electron motion is "one-dimentionalized" and,
as we show below, the Cooper instability restores superconducting
phase. [Note that the suggested above localization of the Q1D
electrons (2) on conducting chains is completely different from
another possible phenomenon - electron localization in
unreasonable high magnetic fields, which correspond to a flux
quantum per unit cell.]

\begin{figure}[h]
\centering
\includegraphics[width=85mm, height=105mm]{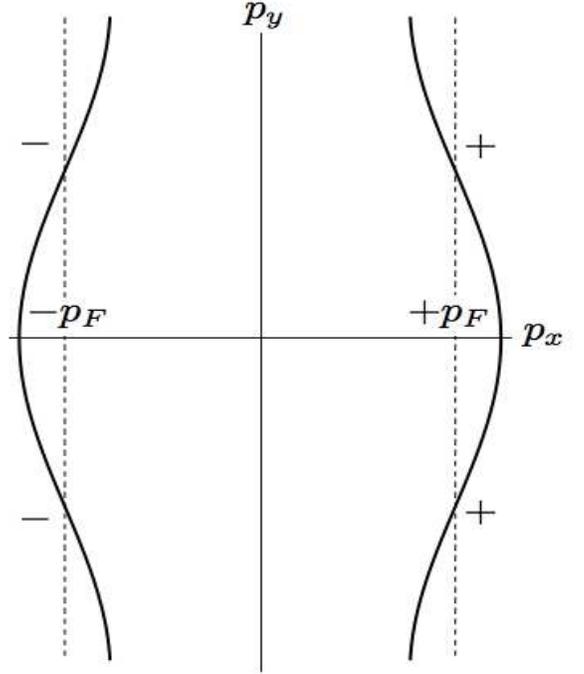}
\caption{Q1D Fermi surface consists of two slightly corrugated sheets extending
in the $z$-direction. The triplet superconducting order parameter changes its
sign on the two sheets of the Q1D Fermi surface.}
\end{figure}

Below, we study the QL superconductivity phenomenon by means of
quantitative quantum mechanical methods appropriate for the
problem under consideration. In a particular, in the magnetic
field (3), we use the Peierls substitution method [5], based on
the Fermi liquid description of Q1D electrons (2):
\begin{eqnarray}
p_x \mp p_F \rightarrow \mp i(d/dx), \ p_y a_y \rightarrow p_y a_y
- \omega_y(\alpha)/v_F,
\nonumber\\
p_z a_z \rightarrow p_z a_z + \omega_z(\alpha)/v_F .
\end{eqnarray}
 As a result, the Schr\"{o}dinger-like equation for electron wave
 functions in the mixed $(p_y,p_z,x)$-representation can be written as:
\begin{eqnarray}
\biggl\{ \mp i v_F \frac{d}{dx} - 2t_y \cos
\biggl[p_ya_y-\frac{\omega_y(\alpha)}{v_F}x \biggl] - 2t_z \cos
\biggl[p_za_z
\nonumber\\
+\frac{\omega_z(\alpha)}{v_F}x \biggl] \biggl\}
\psi_{\epsilon}^{\pm}(p_y,p_z,x) = \delta \epsilon \
\psi_{\epsilon}^{\pm}(p_y,p_z,x),
\end{eqnarray}
where electron energy is counted from the Fermi energy, $\delta
\epsilon = \epsilon - \epsilon_F$, $\epsilon_F = p_F v_F$. Note
that in Eq.(11) we disregard electron spin since we consider below
such triplet superconducting phase where the Pauli paramagnetic
effects do not reveal themselves. It is important that Eq.(11) can
be solved analytically:
\begin{eqnarray}
\psi_{\epsilon}^{\pm}(p_y,p_z,x) =  \exp \biggl( \frac{\pm i
\delta \epsilon x}{v_F} \bigg) \exp  \biggl\{ \pm 2i l_y(\alpha)
\biggl( \sin \biggl[ p_y a_y
\nonumber\\
- \frac{\omega_y(\alpha)}{v_F} x \biggl] -\sin[p_y a_y] \biggl)
\biggl\} \exp  \biggl\{ \mp 2i l_z(\alpha) \biggl( \sin \biggl[
p_z a_z
\nonumber\\
 - \frac{\omega_z(\alpha)}{v_F} x \biggl]
-\sin[p_z a_z] \biggl) \biggl\}.
\end{eqnarray}
Since electron wave functions are known (12), we can define the
finite temperatures Green's functions by means of the standard
procedure [32]:
\begin{eqnarray}
g_{i \omega_n}^{\pm}(p_y,p_z;x,x_1)= \int^{+\infty}_{-\infty}
d(\delta \epsilon) [\psi^{\pm}_{\epsilon}(p_y,p_z;x)]^*
\nonumber\\
\times \psi^{\pm}_{\epsilon}(p_y,p_z;x_1) / (i \omega_n - \delta
\epsilon),
\end{eqnarray}
where $\omega_n$ is the so-called Matsubara frequency.

In the Letter, we consider the following gapless triplet
superconducting order parameter in the Li$_{0.9}$Mo$_6$O$_{17}$,
which, as shown in Refs.[28,29], well satisfies the experimental
data [27]:
\begin{equation}
\hat \Delta(p_x,x) = \hat I  \ sgn(p_x) \ \Delta(x) ,
\end{equation}
where $\hat I$ is a unit matrix in spin-space, $sgn(p_x)$ changes
sign of the triplet superconducting order parameter on two slightly
corrugated sheets of the Q1D FS (2) (see Fig.1).

\begin{figure}[h]
\centering
\includegraphics[width=85mm, height=95mm]{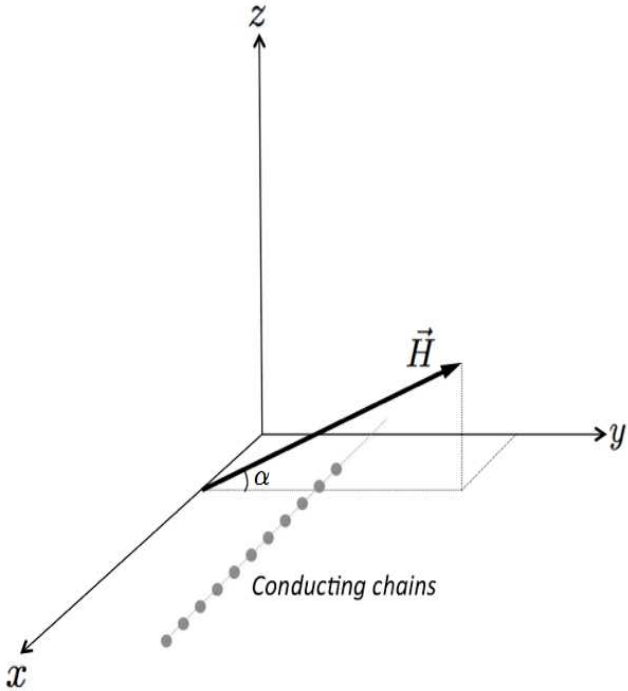}
\caption{The magnetic field makes an angle $\alpha$ with respect to the $y$ axis, perpendicular to the conducting chains.}
\end{figure}

We use the Gor'kov's equations for unconventional superconductivity [33,34]
to obtain the so-called gap equation for superconducting order
parameter $\Delta(x)$. As a result, we derive the following
equation:

\begin{eqnarray}
\Delta(x) = \frac{g}{2}   \int_{|x-x_1| > d} \frac{2 \pi T
dx_1}{v_F \sinh \bigl[ \frac{2 \pi T |x-x_1|}{v_F} \bigl]} \
\Delta (x_1)
\nonumber\\
\times J_0 \biggl\{ 4 l_y(\alpha) \sin \biggl[
\frac{\omega_y(\alpha) (x-x_1)}{2v_F} \bigg] \sin \biggl[
\frac{\omega_y(\alpha) (x+x_1)}{2v_F} \bigg] \biggl\}
\nonumber\\
\times J_0 \biggl\{ 4 l_z(\alpha) \sin \biggl[
\frac{\omega_z(\alpha) (x-x_1)}{2v_F} \bigg] \sin \biggl[
\frac{\omega_z(\alpha) (x+x_1)}{2v_F} \bigg] \biggl\} ,
\end{eqnarray}
where $g$ is the electron coupling constant, and $d$ is the cutoff
distance.

Note that the QL superconductivity phenomenon is directly seen
from Eq.(15). Indeed, at high enough magnetic fields (9), the
parameters $l_y(\alpha)$ and $l_z(\alpha)$ become less than 1.
Under this condition, arguments of the Bessel functions in Eq.(15)
go to zero and superconducting transition temperature goes to its
value in the absence of magnetic field, $T_c$:
\begin{equation}
\lim_{H \rightarrow \infty} T^*_c(H) \rightarrow T_c .
\end{equation}
It is also important that Eq.(15) predicts that superconductivity
in a triplet Q1D superconductor without impurities can survive at
any magnetic fields, including magnetic fields lower than that in
Eq.(9). Nevertheless, we point out that for magnetic fields less
than (9) superconducting transition temperatures are very low and
an account of small amount of impurities would presumably kill the
superconducting phase.

For experimental applications of our results, it is important to
calculate how quickly superconductivity tends to $T_c(0)$ in Eq.(16).
For this purpose, we expand each Bessel function in Eq.(15) to the
second order with respect to the small parameters $l_y(\alpha),
l_z(\alpha) \ll 1$:
\begin{eqnarray}
&&J_0\{...\} \ J_0\{...\} \simeq 1
\nonumber\\
&&- 4 l^2_y (\alpha) \sin^2 \biggl[ \frac{\omega_y(\alpha)
(x-x_1)}{2v_F} \bigg] \sin \biggl[ \frac{\omega_y(\alpha)
(x+x_1)}{2v_F} \bigg]
\nonumber\\
&&- 4 l^2_z (\alpha) \sin^2 \biggl[ \frac{\omega_z(\alpha)
(x-x_1)}{2v_F} \bigg] \sin \biggl[ \frac{\omega_z(\alpha)
(x+x_1)}{2v_F} \bigg] \ .
\end{eqnarray}
In the second approximation with respect to the small parameters
$l_y(\alpha)$ and $l_z(\alpha)$ it is possible to use in the
integral gap equation (15),(17) the following trial function:
\begin{equation}
\Delta(x) = \Delta(y) = const .
\end{equation}
In this approximation, we can also average Eq.(15),(17) over variable
$x+x_1$ and after using the following formula,
\begin{equation}
\frac{1}{g} = \int_d^{+\infty} \frac{2 \pi T_c dz}{\sinh
\biggl(\frac{2 \pi T_c z}{v_F} \biggl)} ,
\end{equation}
we obtain:
\begin{equation}
T_c^*(H)= T_c \biggl\{ 1 - l^2_y(\alpha) \ln \biggl[ \frac{\gamma
\omega_y(\alpha)}{\pi T_c} \biggl] - l^2_z(\alpha) \ln \biggl[
\frac{\gamma \omega_z(\alpha)}{\pi T_c} \biggl]\biggl\},
\end{equation}
where $\gamma$ is the Euler constant.
From Eq.(20) it is directly seen that, at high enough magnetic
fields (9), Eq.(16) is valid.

\begin{figure}[h]
\centering
\includegraphics[width=80mm, height=70mm]{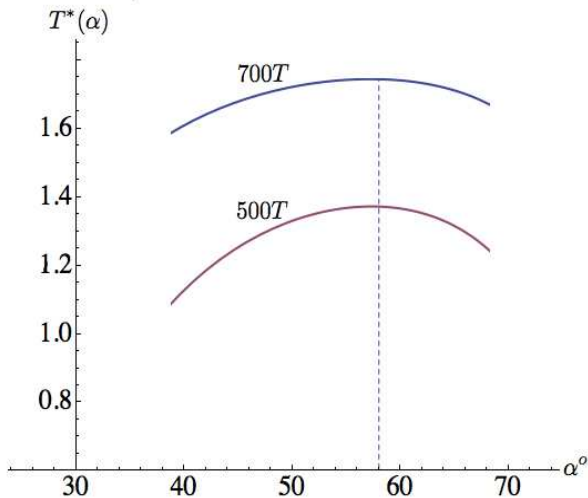}
\caption{The QL superconducting transition temperature dependence
of the angle, $\alpha$, for magnetic fields of $500T$ and $700T$.
Both maximums occur at an angle of $\alpha\approx 58^o$. The
maximum temperature at $700T$ is approximately $1.7K$.}
\end{figure}

For experimental applications of our results it is important to
estimate the value of $T_c^*(H)$ in Eq.(20) for the presumably
triplet [27-29] Q1D superconductor Li$_{0.9}$Mo$_6$O$_{17}$ in
experimental range of magnetic fields. Using known values of the
parameters $a_y$, $a_z$, $v_F$, $t_y$, and $t_z$ (see Table 1 in
Ref.[28]), it is possible to find that
\begin{eqnarray}
\omega_y(H=1 \ T,\alpha=\pi/2)=0.76 \ K,  \nonumber\\
\omega_z(H=1 \ T,\alpha=0)=0.57 \ K, \nonumber\\
l_y(H=1 \ T, \alpha=\pi/2)=116, \nonumber\\
l_z(H=1 \ T, \alpha=0)=49 \ .
\end{eqnarray}

Our next step is to input these parameters into Eq.(20) and
to plot $T_c^*(\alpha)$ as a function of $\alpha$ at given $H$. [We
recall that superconducting transition temperature in the absence
of a magnetic field is equal to $T_c=2.2 \ K$.] In Fig.3, we plot
angular dependence of the QL superconducting phase transition
temperature for two values of a magnetic field, $H = 500 \ T$ and $H
= 700 \ T$. As seen from Fig.3, the maximum values of $T^*_c(H)$
corresponds to angle $\alpha^* \simeq 58^0$ in both cases with
highest $T^*_c(H = 700 \ T) \simeq 1.7 \ K$. Note that region of
validity of Eq.(20) corresponds to the condition $|T^*_c(H)-T_c|
\ll T_c$, therefore, we conclude that Fig.3 correctly represents
calculated transition temperature near its maximums for both values
of the magnetic field.
Note that magnetic fields of the order of $H= 500, \ 700
\ T$ are currently experimentally available as distructive pulsed
magnetic fields.

In the Letter, we have demonstrated that superconductivity can be
restored in a triplet Q1D superconductor in a magnetic field,
perpendicular to its conducting axis, as the Quantum Limit (QL)
superconducting phase. It happens if a magnetic field is high
enough [see Eq.(9)] to localize electrons on conducting chains of
a Q1D conductor. Note that such "one-dimensionalization" of Q1D
electron spectrum promotes also the FISDW instability [5-7] and
non-Fermi-liquid properties [3]. Therefore, we suggest that
superconducting instability is a leading one and that electron
wave function delocalizations between adjacent chains are high
enough for the Fermi-liquid picture to survive. Note that the
FICDW instability [8,9] is not expected in high magnetic fields
since the Pauli paramagnetic effects significantly decrease the
FICDW transition temperature [35]. We suggest to carry out the
corresponding experiment on the presumably triplet superconductor
Li$_{0.9}$Mo$_6$O$_{17}$ in feasibly available pulsed magnetic
fields of the order of $H = 500 - 700 \ T$ and temperatures less
than $T^*_c \simeq 1.4-1.7 \ K$. We have also determined the most
convenient geometry of the experiment which, as shown, corresponds
to inclination angle of $\alpha = 58^0$ [see Eq.(3) and Fig.2]. It
is important that the QL superconductivity phenomenon is not very
sensitive to possible deviations of geometry from the most
convenient one, as seen from Eq.(20) and Fig.3. If the result of
the suggested experiment is positive it will confirm triplet
superconductivity scenario [27-29] in the above mentioned compound
and for the first time establish surviving of superconductivity in
ultra-high magnetic fields.

One of us (A.G.L.) is thankful to N.N. Bagmet and N.E. Hussey for useful
discussions. This work was supported by the NSF under Grant No
DMR-1104512.

$^*$Also at: L.D. Landau Institute for Theoretical Physics, RAS, 2
Kosygina Street, Moscow 117334, Russia.

\end{document}